\documentstyle[twoside,fleqn,espcrc2,epsfig]{article}

\newcommand{\AmS}{{\protect\the\textfont2
  A\kern-.1667em\lower.5ex\hbox{M}\kern-.125emS}}
\def \beq{\begin{equation}}
\def \eeq{\end{equation}}
\def \beqa{\begin{eqnarray}}
\def \eeqa{\end{eqnarray}}
\def \ppbar{\langle\overline\psi\psi\rangle}

\title{Quenched QCD at finite temperature with overlap Fermions
\thanks{This work was funded by the IFCPAR as its project 2104-2.}
}

\author{R.\ V.\ Gavai, Sourendu Gupta\\
   Department of Theoretical Physics,
   Tata Institute of Fundamental Research,\\
   Homi Bhabha Road, Mumbai 400005, India.
   \vskip 6pt R.\ Lacaze\\
   Service de Physique Theorique, CEA Saclay,\\
   F-91191 Gif-sur-Yvette Cedex, France.}
 
\begin{document}

\begin{abstract}
We study quenched QCD just above the phase transition temperature using
overlap Fermions. Exact zero modes of the overlap operator are localized.
Chiral symmetry is restored, as indicated by the behavior of the chiral
condensate after subtracting the effects of zero modes.  The vector
and pseudo-scalar screening masses are close to ideal gas values.
\end{abstract}

\maketitle

In QCD for $T>T_c$ screening correlators in the vector, and
axial-vector channels are saturated by nearly non-interacting quark
anti-quark pairs in the medium \cite{mtc,tifr}. On the other hand,
the scalar and pseudo-scalar screening masses show more complicated
behavior--- strong deviations from the ideal Fermi gas, and a strong
temperature dependence.  This puzzling behavior has been seen in quenched
\cite{quenched} and dynamical simulations with two \cite{nf2} and four
flavors \cite{mtc,tifr,nf4} of staggered quarks, as well as with
Wilson quarks \cite{wilson} close to $T_c$.  The new technique we bring
to bear on this problem is the use of overlap Fermions \cite{neu}. It
has the advantage of preserving chiral symmetry on the lattice for any
number of massless flavors of quarks \cite{luescher}, and hence having
the correct number of pions.

The overlap Dirac operator ($D$) can be defined \cite{neu2} in terms of the
negative Wilson-Dirac operator ($D_w$) by the relation
\beq
   D = 1 - D_w (D_w^\dag D_w)^{-1/2}.
\label{overlap}\eeq
The computation of $D^{-1}$ needs a nested series of two matrix inversions
for its evaluation; each step in the numerical inversion of $D$ involves
the inversion of $D_w^\dag D_w$. Details of the computation of the
inverse square root of the matrix can be found in \cite{mainpaper}.
The study was performed on quenched QCD configurations at temperatures
of $T/T_c$ = 1.25, 1.5 and 2 on $4 \times 8^3$ and $4 \times 12^3$
lattices.  The corresponding couplings are respectively $\beta= $ 5.8,
5.8941 and 6.0625.

A massive overlap operator is defined by
\beq
   D(ma) = ma + (1-ma/2) D = G^{-1}(ma),
\label{moverlap}\eeq
where $m$ is the bare quark mass, $a$ the lattice spacing, and $D$ is
defined in (\ref{overlap}).  We computed the quark propagator, $G$,
on 12 point sources (3 colors and 4 spins) for 10 quark masses from
$ma$=0.001 to 0.5 using a multimass inversion of $D^{\dagger}D$. The
Wilson mass term in $D_w$, which is an irrelevant regulator, was set to
1.8.  The tolerance was $\epsilon=10^{-6}$ in the inner CG and $10^{-4}$
in the outer CG. This meant that the Ginsparg-Wilson relation \cite{gw}
was satisfied to an accuracy of $10^{-9}$, and a chiral Ward identity
to an accuracy of $10^{-5}$.

\begin{figure}[htb]\begin{center}
   \epsfig{file=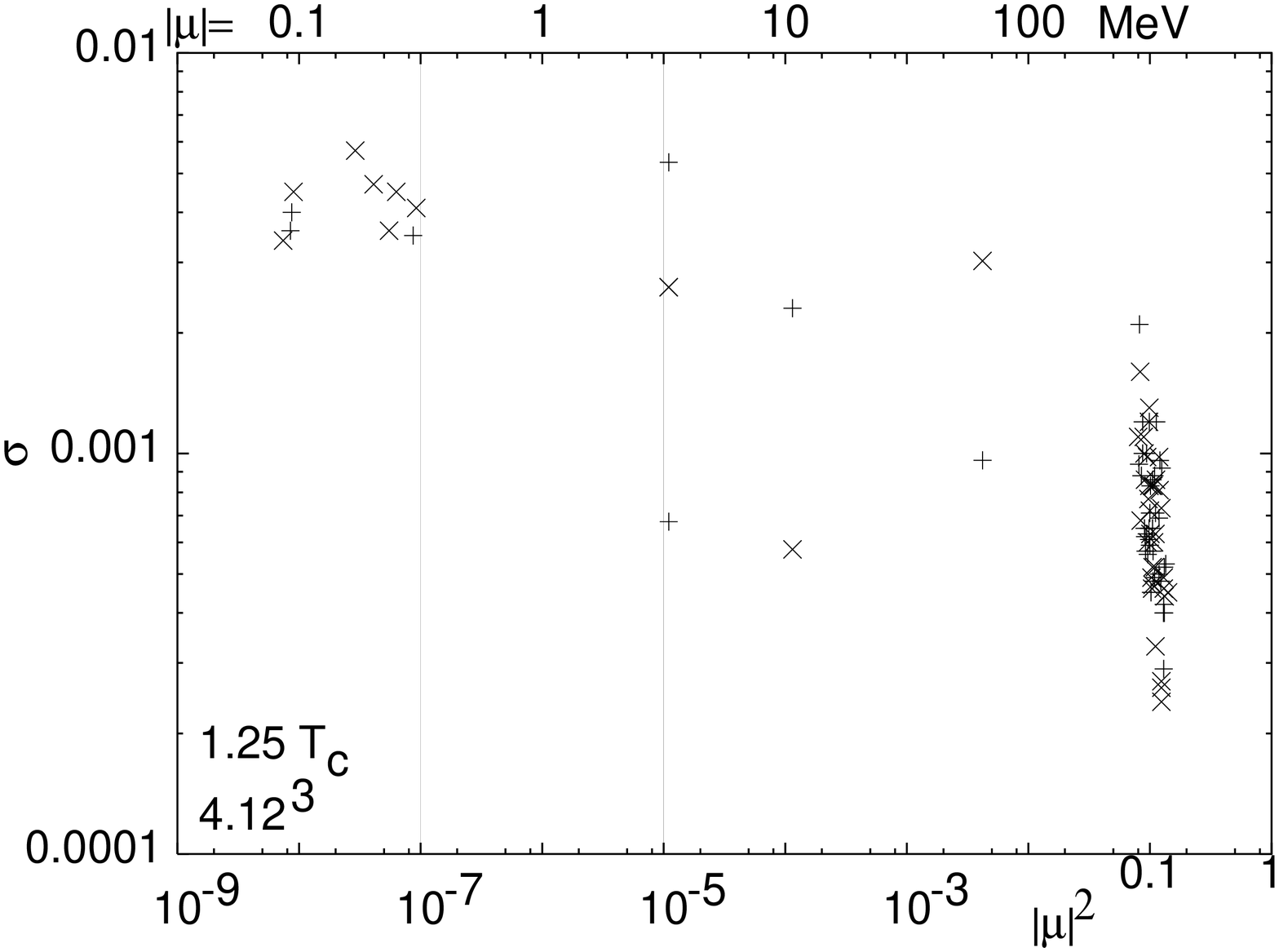,height=5.2cm,width=6cm}
   \end{center}
   \caption{Scatter plot of the eigenvalue, $\mu^2$, and localisation,
    $\sigma$ for the smallest eigenvalues on $4\times12^3$ lattices at
    $T=1.25T_c$. The pluses and crosses denote chiral positive and
    negative sectors respectively.}
\label{fg.spec}\end{figure}

Seperately, a rough computation of the eigenvalues of $D^{\dagger}D$,
$\mu^2$, was made on each configuration with a Lancz\"os method in each
chiral sector.  Whenever $\mu^2 \simeq O(10^{-5})$ was obtained, the
few lowest eigenvalues and eigenvectors were refined to a precision of
about $10^{-8}$ by a Ritz functional minimization.  As shown in Figure
\ref{fg.spec}, for most configurations we found that the lowest $\mu^2$
was well away from zero. However, for some configurations we found zero
and near-zero modes with $\mu^2<10^{-4}$ clearly seperated by a gap
from the non-zero modes with $0.1<\mu^2$.  The non-zero modes clearly
came in degenerate pairs of opposite chiralities.  The zero modes were
all less than $10^{-7}$ and of definite chirality.  There seems to be
a spectral gap between these and the near zero modes (which were seen
only at $T=1.25T_c$). It remains to be seen whether the gap scales with
lattice volume.

We constructed a gauge-invariant measure of localization, $\sigma$
\cite{gockeler}, which varies from unity for an eigenmode localized at
just one site to $1/V$ for an eigenmode spread uniformly on a lattice
of volume $V$.  As shown in Figure \ref{fg.spec}, zero and near-zero
modes tend to be more localized than the non-zero modes.

Our measurement of $\ppbar$ comes from the diagonal part of $G(ma)$
on the 12 sources used for each configuration.  Writing $G$ in terms of
the eigenvectors $\Phi^\mu_\alpha$ of $D^\dag D$ with eigenvalue $\mu^2$
and chirality $\alpha$, the contribution of a zero mode can be easily
read off from the equation above and is seen to be proportional to
$\Phi\overline\Phi/ma$. Since the eigenvector corresponding to the zero
mode is localized, its contribution to the condensate depends strongly
on the spatial position of the source vector. The remaining modes are
delocalized and closely spaced; so the overlap of the eigenvector on the
source is averaged out.  After subtracting the zero mode contribution,
$\ppbar$ is strikingly identical to that seen in the sample without
zero modes, at all the couplings and lattice sizes studied.  We found
that $\ppbar$ varies linearly with $ma$ and goes to zero as $ma\to0$,
with a value of $\ppbar/ma$ which is independent of the lattice volume.
This is how chiral symmetry restoration manifests itself in quenched QCD.

\begin{figure}[htb]\begin{center}
   \epsfig{file=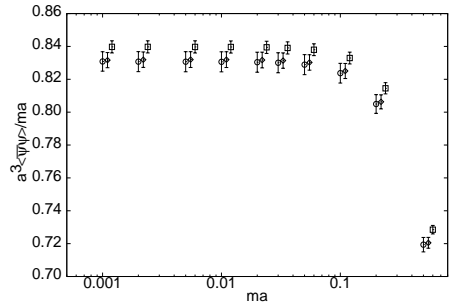,height=6cm,width=4.2cm,angle=-90}
   \end{center}
   \caption{$a^3\ppbar/ma$ as a function of $ma$ at $T=1.25T_c$ (circle),
      $1.5T_c$ (diamond) and $2T_c$ (square) on $4\times12^3$ lattices. 
      Zero and near-zero mode contributions are subtracted. Data for
      $1.5T_c$ and $2T_c$ are displaced in $ma$ for visibility.}
\label{fg.cond}\end{figure}

\begin{figure}[htb]\begin{center}
    \epsfig{file=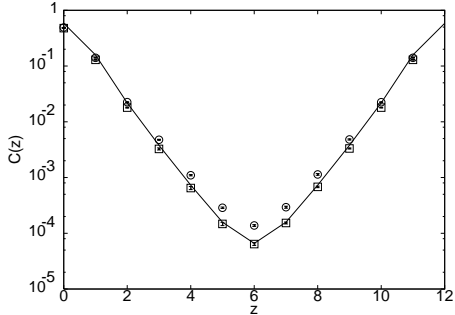,height=4.2cm,width=6cm}
   \end{center}
   \caption{The screening correlators on $4.12^3$ lattices at
      $T=1.5T_c$ for $ma=0.001$, in analyses without zero modes. The
      V/AV correlators (boxes) agree very well with the ideal gas
      computation (solid line), and the S/PS correlators (circles)
      are also similar.}
\label{fg.ideal}\end{figure}

The following identities hold for overlap Fermions
in the chirally symmetric phase as $ma\to0$,
\beq
   C_S(z) = -C_{PS}(z) \quad{\rm\ and}\quad C_V(z)=C_{AV}(z).
\label{corid}\eeq
Here $C$ is the screening correlation function in the spatial
$z$ direction of an operator summed in the other three directions.
The subscripts PS refer to a pseudo-scalar operator,
S to a scalar, V to a vector and AV to an axial-vector.
We find that the V and AV correlators indeed agree at all
$z$ and at all temperatures we studied. The S and PS obey this
relation after the zero modes have been subtracted. In addition,
$C_V$ is described well by an ideal gas of overlap
quarks on the same lattice, while $C_{PS}$ gives a screening mass
within 10\% of the ideal has result (see Figure \ref{fg.ideal}
for an example).

In conclusion, working with chiral (overlap) Fermions, we have found
several new results and a consistent picture of the high temperature phase
of quenched QCD.  The quenched thermal ensemble contains gauge fields
which give rise to Fermion zero modes of definite chirality. When the
effect of these modes is subtracted, $\ppbar$ vanishes in the zero quark
mass limit, showing that chiral symmetry is restored. Simultaneously,
parity doubling is seen in the spectrum of screening masses, which are
close to those expected in an ideal Fermi gas, even for the S/PS sector.
Since some of these results are not obtained with staggered quarks, it
is an interesting question whether the two flavor QCD phase transition
is properly described by such a representation of quarks.

Some interesting problems remain to be solved.  At $T\le1.25\ T_c$, there
are near-zero modes.  It cannot be ruled out that these modes shift the
quenched chiral symmetry restoration point away from $T_c$.  However,
this question is crucially related to the evolution of near-zero modes
with lattice volume and spacing. Hence the nature of these complications,
and the question of whether they are quenched artifacts or remain in
full QCD, will only become clear with further studies which are underway.

\end{document}